\documentclass{iopart}
\usepackage{iopams} 
\usepackage{epsfig}  
\usepackage{graphics}  

\newcommand{\erfc}      {{\rm erfc}}
\newcommand{\BetaB}     {{\rm B}}
\newcommand{\ParabolicD}{{\rm D}}
\newcommand{\MoshinskyM}{{\rm M}}
\newcommand{\ch}        {{\rm ch}}

\begin{document}

\title{Tunneling out of a time-dependent well}

\author{Tobias Kramer\footnote[1]
{Present address: 
Physik-Department T30,
Technische Universit\"at M\"unchen,
James-Franck-Stra{\ss}e,
85747 Garching, Germany} 
and Marcos Moshinsky\footnote[2]{Member of El Colegio Nacional and Sistema Nacional de Investigadores}}

\address{
Instituto de F\'{\i}sica,
Universidad Nacional Aut\'onoma de M\'exico,
Apartado Postal  20-364,
01000 M\'exico D.F., M\'exico}

\ead{tkramer@ph.tum.de, moshi@fisica.unam.mx}

\begin{abstract}

Solutions to explicit time-dependent problems in quantum mechanics are rare. In fact, all known solutions are coupled to specific properties of the Hamiltonian and may be divided into two categories: One class consists of time-dependent Hamiltonians which are not higher than quadratic in the position operator, like i.e.\ the driven harmonic oscillator with time-dependent frequency. The second class is related to the existence of additional invariants in the Hamiltonian, which can be used to map the solution of the time-dependent problem to that of a related time-independent one.

In this article we discuss and develop analytic methods for solving time-dependent tunneling problems, which cannot be addressed by using quadratic Hamiltonians. Specifically, we give an analytic solution to the problem of tunneling from an attractive time-dependent potential which is embedded in a long-range repulsive potential.

Recent progress in atomic physics makes it possible to observe experimentally time-dependent phenomena and record the probability distribution over a long range of time. Of special interest is the observation of macroscopical quantum-tunneling phenomena in Bose-Einstein condensates with time-dependent trapping potentials. We apply our model to such a case in the last section.
\end{abstract}

\submitto{\JPA}

 \maketitle
 
\section{Solutions to time-dependent problems in quantum-mechanics}

In this paper we construct the time-evolution operator (or propagator) $K(x,t|x',t')$ for some (time-dependent) potentials in position space. The use of the propagator enables us to discuss the time-evolution of an initial state $\psi(x,t')$, which is given at time $t'$:
\begin{equation}\label{eq:TimeEvo}
\psi(x,t)=\int\rmd x'\,K(x,t|x',t')\,\psi(x',t').
\end{equation}
Surprisingly, it is very hard to find analytic expressions for the propagator (for a comprehensive list see \cite{Grosche1998a}). Time-dependent problems in quantum-mechanics are divided into two groups:
\begin{enumerate}
\item The sudden switch from one Hamiltonian to another one at time $t'$, where both Hamiltonians are not time-dependent:
\begin{equation}
H=
\left\{
\begin{array}{ll}
H_-&{\rm if}\;t<t'\\
H_+&{\rm if}\;t\ge t'
\end{array}\right. .
\end{equation}
The sudden change of the potential will lead to a spatial time-evolution of the wave function $\psi(x,t)$, since in general a stationary eigenstate of $H_-$ will not be a stationary eigenstate of $H_+$. In this sense, $H_-$ generates an initial wavepacket, which will propagate under the influence of $H_+$ for $t>t'$.
\item Explicit time-dependence due to a continuously changing Hamiltonian $H(t)$ for $t\ge 0$.
\end{enumerate}
In the first case, one can expand the given eigenstate of $H_-$ in terms of the eigenstates of $H_+$, or if one models a continuous source of particles (i.e.\ a particle beam) one can use the powerful method of Laplace transforms, either in an approach that directly incorporates the initial wave function $\psi(x,t')$, or by construction of the energy-dependent Green function. Instructive examples are the problem of diffraction in time (the Moshinsky shutter) \cite{Moshinsky1952a} and the evolution of quantum states in external fields \cite{Kleber1994a,Kramer2002a}.

\begin{table}[t]
\begin{center}
\begin{tabular}{|l|l|}
\hline Hamiltonian $H(t)$ & Reference \\ 
\hline $H(t)=\frac{p^2}{2m}-V_0\delta(x)/t$ & \cite{Kleber1994a} \\ 
\hline $H(t)=\frac{p^2}{2m}-V_0\delta(x)/\sqrt{t^2+B t+C}$ & \cite{Dodonov1992a} \\
\hline $H(t)=\frac{p^2}{2m}-\frac{1}{2}m\omega^2 x^2-V_0\rme^{-\omega t}\delta(x)$ & this work \\
\hline 
\end{tabular} 
\end{center}
\caption{Known analytic solutions to time-dependent Hamiltonians $H(t)$ that are not quadratic in $x,p$.\label{tab:solutions}}
\end{table}

Unfortunately the method of Laplace transforms does not cope well with time-dependent potentials. For this type of problems, the Laplace transform gives rise to integro-differential equations. From a conceptual point of view this obstacle is not unexpected, since it is no longer possible to work with a conserved energy. Analytic expressions for propagators are only known for two remaining subclasses:
\begin{itemize}
\item[1)] quadratic Hamiltonians with time-dependent coefficients (i.e.\ the driven harmonic oscillator with a time-dependent frequency) \cite{Grosche1998a},
\item[2)] time-dependent potentials that possess a time-dependent scaling parameter (all the potentials in Table~\ref{tab:solutions} belong to this class) \cite{Ray1982a,Dodonov1992a,Kleber1994a}.
\end{itemize}
Of special interest for tunneling phenomena are potentials that possess a dip and a barrier through which an initially confined quantum state can decay (see Fig.~\ref{fig:WellDecay}). Commonly used models for tunneling barriers are cubic, Eckart-type, or $\delta$ potentials \cite{Maitra1997a,Saltzer2003a,Elberfeld1988a,Andreata2004a}.

\section{Tunneling out of a one-dimensional quantum well}

In this work we will derive a novel analytic solution for the tunneling out of a quantum well (modelled by an attractive $\delta$-potential $V(x,t)=-V(t)\delta(x)$) which is embedded into a repelling oscillator potential $H_\cap=\frac{p^2}{2m}-\frac{1}{2}m\omega^2 x^2$. 
We will discuss and compare three different cases:
\begin{enumerate}
\item The propagator for an inverted oscillator potential:
\begin{equation}\label{eq:CaseI}
H=\frac{p^2}{2m}-\frac{1}{2}m\omega^2 x^2.
\end{equation}
\item The propagator for a time-independent attractive $\delta$-potential embedded in a repelling oscillator:
\begin{equation}\label{eq:CaseII}
H=\frac{p^2}{2m}-\frac{1}{2}m\omega^2 x^2-V_0\delta(x).
\end{equation}
\item The propagator for an increasingly less attractive $\delta$-potential embedded in a repelling oscillator:
\begin{equation}\label{eq:CaseIII}
H=\frac{p^2}{2m}-\frac{1}{2}m\omega^2 x^2-V_0\rme^{-\omega t}\delta(x).
\end{equation}
\end{enumerate}

Since we are interested in the decay rate of the initially confined state, we have to define a quantity that will convey this information. One possibility is to calculate the ,,survival probability'', which is the overlap of an initial state with the state at a later time:
\begin{equation}\label{eq:P}
\fl
P(t)={\left|\int\rmd x\,\int\rmd x'\,\psi^*(x,0)K(x,t|x',0)\psi(x',0)\right|}^2
={\left|\int\rmd x\,\psi^*(x,0)\psi(x,t)\right|}^2
\end{equation}
For the initial state, we will always chose the bound eigenstate of the Hamiltonian
$H_\delta=\frac{p^2}{2m}-V_0\delta(x)$:
\begin{equation}\label{eq:PsiDelta}
\psi_{\delta}(x)=\frac{\sqrt{m V_0}}{\hbar}\exp\left[-\frac{m V_0 |x|}{\hbar^2}\right].
\end{equation}

\subsection{Propagation under the influence of an inverted oscillator}

The propagator for
\begin{equation}
H_\cap=\frac{p^2}{2m}-\frac{1}{2}m\omega^2 x^2
\end{equation}
is given by setting $\omega=\rmi\omega_{\rm at}$ in the propagator for the attractive oscillator \cite{Kennard1927a,Grosche1998a}:
\begin{equation}\label{eq:KISHO}
\fl
K_\cap(x,t|x',0)
=\sqrt{\frac{m\omega}{2\pi\rmi\hbar\sinh(\omega t)}}
\exp\left[
\frac{\rmi m \omega}{\hbar}\frac{(x^2+x^{\prime 2})\cosh(\omega t)-2xx'}{2\sinh(\omega t)}
\right]
\end{equation}
The time evolution of the initial state $\psi_{\delta}(x)$ (\ref{eq:PsiDelta}) can be expressed in terms of the Moshinsky function $\MoshinskyM(x,k,t)$ \cite{Moshinsky1952a, Nussenzveig1992a}, which may be defined in different ways:
\begin{eqnarray}\label{eq:DefM}
\MoshinskyM(x,k,t)
&=\int_{-\infty}^0\rmd x'\frac{1}{\sqrt{2\pi\rmi t}}\exp\left[\rmi k x'+\rmi\frac{{(x-x')}^2}{2t}\right]\\
&=\frac{\rmi}{2\pi}\int_{-\infty}^\infty\rmd \kappa 
\frac{\exp\left[\rmi\kappa x-\frac{\rmi}{2}\kappa^2 t\right]}{\kappa-k}\\
&=\frac{1}{2}\exp\left[\rmi k x-\frac{\rmi}{2}k^2 t\right]
\erfc\left[\rme^{-\rmi\pi/4}\frac{(x-kt)}{\sqrt{2t}}\right].
\end{eqnarray}
Using equations~(\ref{eq:TimeEvo}) and (\ref{eq:DefM}), we obtain the wave function at a later time $t$:
\begin{eqnarray}
\fl
\psi_\cap(x,t)=\sqrt{\frac{m^2\omega V_0}{2\pi\rmi\hbar^3\sinh(\omega t)}}
\int_{-\infty}^\infty\!\!\!\rmd x'
\exp\left[
\frac{\rmi m \omega}{\hbar}\frac{(x^2+x^{\prime 2})\cosh(\omega t)-2xx'}{2\sinh(\omega t)}
-\frac{m V_0 |x'|}{\hbar^2}\right]
\nonumber\\
\fl\quad=\frac{1}{\hbar}\sqrt{\frac{m V_0}{\coth(\omega t)}}
\left[
\MoshinskyM\bigg(x,
-\frac{\rmi m V_0}{\hbar^2}
+\frac{x m\omega}{\hbar}\left(\frac{1}{\tanh(\omega t)}-\frac{1}{\sinh(\omega t)}\right),
\frac{\hbar\tanh(\omega t)}{m\omega}\right)
\nonumber\\
\fl\quad+\MoshinskyM\left(-x,
-\frac{\rmi m V_0}{\hbar^2}
-\frac{x m\omega}{\hbar}\left(\frac{1}{\tanh(\omega t)}-\frac{1}{\sinh(\omega t)}\right),
\frac{\hbar\tanh(\omega t)}{m\omega}\bigg)
\right]
\end{eqnarray}
\begin{figure}[t]
\begin{center}
\includegraphics[width=0.65\textwidth]{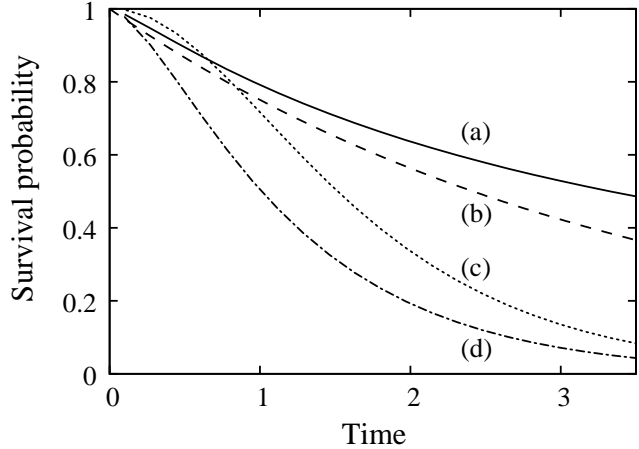}\\
\end{center}
\caption{\label{fig:DecayISHO}Survival probability of the initial state $\psi_\delta(x)=\rme^{-|x|}$ for different potentials: (a) free particle, (b) inverted oscillator and static $\delta$-potential, (c) inverted oscillator and exponentially decreasing $\delta$-potential, (d) inverted oscillator potential. Parameters: atomic units ($m=\hbar=1$), $\omega=1$, $V_0=1$.}
\end{figure}
The corresponding survival probability is shown in Fig.~\ref{fig:DecayISHO}. Not surprisingly, the presence of a repelling potential expedites the decay process compared to the free evolution (with $\omega\rightarrow 0$) of the wave packet, which is also displayed.

\subsection{Tunneling from a time-independent quantum well}\label{sec:stationary_delta+isho}

A more interesting case is the combination of a repelling long-range potential with an attractive short-range potential. A general method to derive the propagator for the combination of a $\delta$-function potential and another potential is based on the solution of the Lippmann-Schwinger equation by using a Laplace transform (for examples and a discussion see \cite{Kleber1994a}, Section~3.4.). This method is well suited for time-independent Hamiltonians of the form
\begin{equation}
H=H_0-V_0\delta(x).
\end{equation}
In this case the Lippmann-Schwinger equation becomes
\begin{eqnarray}\label{eq:LS}
\fl
K(x,t|x',0)=K_0(x,t|x',0)-\frac{\rmi}{\hbar}\int_0^{t}\!\rmd t''\int_{-\infty}^\infty\!\!\!\rmd x'' K_0(x,t|x'',t'')\,V(x'',t'')\,K(x'',t''|x',0)\nonumber\\
\lo=
K_0(x,t|x',0)
+\frac{\rmi V_0}{\hbar}\int_0^{t}\rmd t''
K_0(x,t|0,t'')\,K(0,t''|x',0),
\end{eqnarray}
where $K_0$ denotes the propagator for the simpler Hamiltonian $H_0$. In order to solve this equation for the propagator $K$ of the full Hamiltonian $H$, it is of advantage to switch to a Laplace transformed picture. The Laplace transform of the propagator is conveniently expressed in terms of the energy-dependent Green function $G(x,x';E)$ \cite{Economou1983a}:
\begin{equation}
G(x,x';E)=\frac{1}{\rmi\hbar}\lim_{\eta\rightarrow 0}\int_0^\infty\rmd t\,
\rme^{\rmi (E+\rmi\eta)t/\hbar}K(x,t|x',0).
\end{equation}
If we apply the Laplace transform to both sides of equation~(\ref{eq:LS}), we can use the convolution theorem for a product of two functions (see \cite{Abramowitz1965a}, 29.2.8) and obtain the Green $G$ function of the full Hamiltonian $H$ in terms of the Green function $G_0$ for $H_0$:
\begin{equation}\label{eq:GLS}
G(x,x';E)=G_0(x,x';E)-V_0\frac{G_0(x,0;E) G_0(0,x';E)}{1+V_0 G_0(0,0;E)}.
\end{equation}
The inverse Laplace-transform yields the propagator:
\begin{equation}\label{eq:KLS}
\fl
K(x,t|x',0)=K_0(x,t|x',0)+\frac{V_0}{2\pi\rmi}
\int_{-\infty}^{\infty}\rmd E\,\rme^{-\rmi E t/\hbar}
\frac{G_0(x,0;E) G_0(0,x';E)}{1+V_0 G_0(0,0;E)}.
\end{equation}
For the case of a $\delta$-function potential on top of an inverted oscillator we have
\begin{equation}
\fl
H=H_0+V(x),\qquad H_0=\frac{p^2}{2m}-\frac{1}{2}m\omega^2x^2,
\qquad{\rm and}\qquad 
V(x)=-V_0\delta(x).
\end{equation}
The Laplace transform of the propagator for the inverted oscillator (\ref{eq:KISHO}) gives rise to the parabolic cylinder function (we use \cite{Gradshteyn1994a}, 3.547(10) [there is a misprint, the correct result is $\int_0^{\infty}\exp \left\{-2(\alpha \coth x + p x)\right\} \sinh^{2\beta}x\,dx=\frac{1}{2}\alpha^{\beta}\Gamma (p-\beta)
W_{-p,\beta+\frac{1}{2}}(4\alpha)$] and 9.240):
      
\begin{eqnarray}\label{eq:KISHOLaplace}
\fl G_0(x,0;E)=\frac{1}{\rmi\hbar}\int_0^\infty\rmd t\,\rme^{\rmi E t/\hbar}K_\cap(x,t|0,0)\nonumber\\
\lo=\frac{2^{-3/4+E/(2\rmi\hbar\omega)}}{\rmi\hbar} \sqrt{\frac{m}{\rmi\pi\hbar\omega}}
\Gamma\left(\frac{1}{4}+\frac{E}{2\rmi\hbar\omega}\right)
\ParabolicD_{-\frac{1}{2}-\frac{E}{\rmi\hbar\omega}}
\left(\sqrt{2}|x|\sqrt{\frac{m\omega}{\rmi\hbar}}\right).
\end{eqnarray}
At the origin this expressions simplifies to
\begin{equation}
G_0(0,0;E)=\frac{1}{2\rmi\hbar\omega}\sqrt{\frac{m\omega}{\pi\rmi\hbar}}
\BetaB\left[\frac{E}{2\rmi\hbar\omega}+\frac{1}{4},\frac{1}{2}\right],
\end{equation}
where $\BetaB[x,y]$ denotes Euler's beta function (see \cite{Abramowitz1965a}, 6.2).
In principle, we could now carry out the inverse Laplace transform by integrating along the real energy axis. But since our main interest is to determine the tunneling rate for this type of potential, another method is available to extract this information. Following the procedure suggested by Ludviksson for the combination of a $\delta$-Potential with a linear force field \cite{Ludviksson1987a}, we locate the poles of the denominator in equation~(\ref{eq:GLS}) in the complex energy plane. For the inverted oscillator, they are given by the roots of the transcendental equation
\begin{equation}
1+V_0 G_0(0,0;E)=0.
\end{equation}
The determination of the roots (denoted by $E_\delta^{(i)}$) has to be done numerically. Note, that there is another set of poles $E_\Gamma$ present, which results from the $\Gamma$-function in $G_0(x,0;E)$:
\begin{equation}
E_\Gamma^{(k)}=-\rmi\hbar\omega\left(k+\frac{1}{2}\right), \quad k=0,2,4,\ldots.
\end{equation}
All poles have a negative imaginary part and thus lead to a decaying wave-function for $t\rightarrow\infty$ and fixed locations $x,x'$. However, none of the poles should be included in an evaluation of the inverse Laplace transform, since the original contour runs along the real energy axis and cannot be closed in the lower imaginary $E$-plane: The residues at the poles with negative imaginary part correspond to wavefunctions that are not bounded at infinity (see also the discussion in \cite{Ludviksson1987a}). However, the first pole $E_\delta^{(1)}$ lies closest to the real energy axis and is directly connected to the presence of the $\delta$-Potential:
For $\omega\rightarrow 0$ the pole is located at $E=-m V_0^2/(2\hbar^2)$, which is precisely the binding energy of the $\delta$-Potential. The inverted-oscillator potential leads to a shift of this pole to
\begin{equation}
E_\delta^{(1)}=E_r-\frac{\rmi}{2}E_i,\quad E_r,E_i\in \mathbf{R}.
\end{equation}
An estimate for the lifetime of the decaying state is thus given by
\begin{equation}
T_{\rm life}=\frac{\hbar}{E_i}.
\end{equation}
Fig.~\ref{fig:DecayISHO} shows that the exponential decay rate is slowed down compared to the evolution in a purely inverted oscillator potential due to the attraction of the $\delta$-potential.

\subsection{Time-dependent $\delta$-potential and an inverted oscillator}\label{sec:Mapping}

Interestingly, it is possible to obtain an exact analytic expression for the propagator of a time-dependent $\delta$-potential on top of an inverted harmonic oscillator. The basic idea is to employ a mapping procedure, which transforms the explicit time-dependent problem into a related time-independent one. Details of this method are given in \cite{Ray1982a,Dodonov1992a,Grosche1998a}. The potential must have the following explicit time-dependence:
\begin{equation}\label{eq:VDGL}
V(x,t)=\frac{1}{C^2(t)}V\left(\frac{x}{C(t)}+A(t)\right)+g_1(t) x+g_2(t) x^2+g_0(t),
\end{equation}
where the time-dependent coefficients $C(t),A(t),g_0(t),g_1(t),g_2(t)$ are coupled by a set of ordinary differential equations:
\begin{eqnarray}
\ddot{C}(t)-2 g_2(t) C(t)=K/C^3(t),\\
\ddot{A}(t)+\frac{2\dot{A}(t)\dot{C}(t)}{C(t)}+\frac{K A(t)}{C^4(t)}-\frac{g_1(t)}{C(t)}=0,
\end{eqnarray}
and where $K$ denotes an arbitrary constant. If we choose for the potential the form
\begin{equation}\label{eq:VDeltaISHOTime}
V(x,t)=-\frac{1}{2}m\omega^2x^2-\rme^{-\omega t} V_0 \delta(x),
\end{equation}
we can fulfill all the differential equations by setting
\begin{equation}
\fl
C(t)=\rme^{\omega t},\qquad A(t)=g_0(t)=g_1(t)=K=0,\qquad g_2(t)=-\frac{1}{2}m\omega^2.
\end{equation}
Following \cite{Ray1982a}, we can now construct an equivalent time-independent Hamilton operator with a potential $\tilde{V}(\tilde{x})$ by transforming the position operator and the time according to
\begin{equation}\label{eq:TransformedCoords}
\fl
\tilde{x} =\frac{x}{C(t)}=\rme^{-\omega t}x,\qquad
\tilde{x}'=\frac{x'}{C(0)}=x',\qquad
\tilde{t} =\int_0^t\frac{\rmd s}{C^2(s)}=\frac{1-\rme^{-2\omega t}}{2\omega}.
\end{equation}
The propagator for the transformed Hamiltonian
\begin{equation}
\tilde{H}=\frac{\tilde{p}^2}{2m}+\tilde{V}(\tilde{x})
         =\frac{\tilde{p}^2}{2m}-V_0\delta(\tilde{x})
\end{equation}
can be derived with the method of the previous section with the free particle Green function ($k=\sqrt{2mE}/\hbar$)
\begin{equation}
\fl
G_{\rm free}(x,x';E)=-\frac{\rmi m}{\hbar^2 k}\rme^{\rmi k |x-x'|},
\quad K_{\rm free}(x,t|x',0)=\sqrt{\frac{m}{2\pi\hbar t}}\exp\left[\frac{\rmi m{(x-x')}^2}{2\hbar t}\right]
\end{equation}
The inverse Laplace transform is conveniently expressed in terms of the Moshinsky function (see i.e.\ equation~(3.43) in  \cite{Kleber1994a}, replace $V_0$ by $-V_0$):
\begin{eqnarray}\label{eq:Ktilde}
\fl
\tilde{K}(\tilde{x},\tilde{t}|\tilde{x}',0)
=K_{\rm free}(\tilde{x},\tilde{t}|\tilde{x}',0)+
\frac{m V_0}{2 \pi \rmi\hbar^2}\int_{-\infty}^\infty\rmd k\,
\frac{\exp\left[\frac{-\rmi \hbar k^2 \tilde{t}}{2 m}+\rmi k(|\tilde{x}|+|\tilde{x}'|)\right]}{k-\rmi m V_0/\hbar^2}\\
\lo=K_{\rm free}(\tilde{x},\tilde{t}|\tilde{x}',0)+
\frac{m V_0}{\hbar^2}
\MoshinskyM(|\tilde{x}|+|\tilde{x}'|,+\rmi m V_0/\hbar^2,\hbar \tilde{t}/m).
\end{eqnarray}
The mapping is associated with a canonical transformation that gives rise to additional factors in the propagator \cite{Ray1982a}:
\begin{eqnarray}\label{eq:Kres}
\fl
K(x,t|x',0)=
\frac{1}{\sqrt{C(t)C(0)}}
\exp\left[\frac{\rmi m}{2\hbar}\left(
  x^2         \frac{\dot{C}(t)}{C(t)}
 -x^{\prime 2}\frac{\dot{C}(0)}{C(0)}
\right)\right]
\tilde{K}(\tilde{x},\tilde{t}|\tilde{x}',0)\\
\lo=
\exp\left[-\frac{\omega t}{2}\right]
\exp\left[\frac{\rmi m \omega(x^2-x^{\prime 2})}{2\hbar}\right]
\tilde{K}\left(\rme^{-\omega t}x,\frac{1-\rme^{-2\omega t}}{2\omega}\bigg|x',0\right).
\end{eqnarray}
Inserting equation~(\ref{eq:Ktilde}) shows that the exponential factor $\exp\left[-\frac{\omega t}{2}\right]$ is compensated by $K_{\rm free}(\tilde{x},\tilde{t}|\tilde{x}',0)$, and the free propagator is mapped into the one of the inverted oscillator (as expected from equation~(\ref{eq:VDeltaISHOTime}) in the limit $V_0\rightarrow 0$):
\begin{equation}
\exp\left[-\frac{\omega t}{2}\right]
\exp\left[\frac{\rmi m \omega(x^2-x^{\prime 2})}{2\hbar}\right]
K_{\rm free}(\tilde{x},\tilde{t}|\tilde{x}',0)=K_\cap(x,t|x',0).
\end{equation}
Therefore we may rewrite equation~(\ref{eq:Kres}) like equation~(\ref{eq:KLS}) as the sum of the propagator for the inverted oscillator $K_\cap$ and a contribution due to the time-dependent potential $K_V$:
\begin{equation}
K(x,t|x',0)=K_\cap(x,t|x',0)+K_{V}(x,t|x',0),
\end{equation}
with
\begin{equation}
\fl
K_{V}(x,t|x',0)=
\frac{m V_0}{\hbar^2}
\rme^{\rmi m \omega(x^2-x^{\prime 2})/(2\hbar)-\frac{\omega t}{2}}
\MoshinskyM\left(
 \rme^{-\omega t}|x|+|x'|,
 \frac{\rmi m V_0}{\hbar^2},
 \frac{\hbar (1-\rme^{-2\omega t})}{2\omega m}\right).
\end{equation}
The corresponding survival probability of the initial state (\ref{eq:PsiDelta}) is displayed in Fig.~\ref{fig:DecayISHO}. As expected, the decreasing $\delta$-potential strength expedites the decay process. The asymptotic behaviour for $\omega t\gg 1$ can be extracted by inserting 
$\lim_{\omega t\rightarrow\infty} \tilde{x}=0$, and $\lim_{\omega t\rightarrow\infty} \tilde{t}=1/(2\omega)$ into equation~(\ref{eq:Ktilde}):
\begin{eqnarray}
\fl
K(x,t|x',0)\approx
\exp[-\omega t/2]
\left[\rme^{+\rmi m \omega x^2/(2\hbar)}\right]
\left[\rme^{-\rmi m \omega x'^2/(2\hbar)}\tilde{K}(0,1/(2\omega)|x',0)\right].
\end{eqnarray}
Using equation~(\ref{eq:P}) we obtain an exponentially decaying survival probability
\begin{eqnarray}
\fl
P(t)\approx\rme^{-\omega t}
{\left|\int\rmd x\,\rme^{\rmi m \omega x^2/(2\hbar)}\psi^*(x,0)\right|}^2
{\left|\int\rmd x' \rme^{-\rmi m \omega x'^2/(2\hbar)}\tilde{K}(0,\frac{1}{2\omega}|x',0)\psi(x',0)\right|}^2\!\!\!.
\end{eqnarray}
Note that this behaviour is independent of the choice of $V(x)$ in equation~(\ref{eq:VDGL}).

\section{The optical atom laser as a time-dependent problem}\label{sec:optical_atomlaser}

\begin{figure}[t]
\begin{center}
\includegraphics[width=0.5\textwidth]{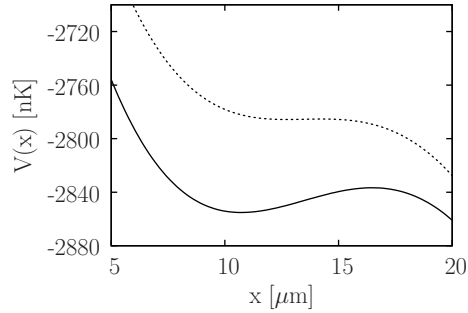}\\
\end{center}
\caption{\label{fig:OpticalPotential}Optical trapping potential for Rubidium atoms in the presence of the gravitational force along the $x$-axis: $V(x)=-V_0\,\exp(-2 x^2/w_0^2)-F x$. Parameter: Lower curve $V_0=2400$~nK, upper one: $V_0=2286$~nK, $w_0=27\mu$m, $F=103$~nK/$\mu$m. Only the lower curve can support a metastable state.}
\end{figure}
As an application of our method we discuss the time evolution of the quantum-state of a Bose-Einstein condensate (BEC) that is confined in a time-dependent trapping potential. Recent experiments use optically trapped BECs of Rubidium atoms to produce a matter wave of macroscopic dimensions \cite{Cennini2003a}. Optical traps utilize the polarizability of atoms to achieve a BEC in the presence of a running optical wave, which is generated by a focussed laser beam. To a good approximation, the effective optical potential of the laser is given by \cite{Cennini2004a}
\begin{equation}
V_{\rm trap}(x,y,z,t)=-\frac{V(t)}{1+{(z/z_0)}^2}\,
\exp\left[-\frac{2(x^2+y^2)}{w_0^2}\frac{1}{1+{(z/z_0)}^2}\right],
\end{equation}
where 
\begin{equation}
z_0=\frac{\pi w_0^2}{\lambda_{\rm laser}}, 
\qquad{\rm and}\qquad
V(t)=\frac{\alpha_s P_{\rm laser}(t)}{\pi\epsilon_0 c w_0^2}.
\end{equation}
Typical experimental parameters are reported in \cite{Cennini2003a}: The laser is characterized by its minimal beam waist $w_0$ (27~$\mu$m), its wavelength $\lambda_{\rm laser}$ (10.6~$\mu$m) and an adjustable power $P_{\rm laser}$ (200-0 mW). The polarizability $\alpha_s$ of $^{87}$Rb atoms is about $5.3\times 10^{-39}$m$^2$C/V. The other constants denote the permittivity of the vacuum $\epsilon_0$ and the speed of light $c$.

Since the laser is operated in the presence of the gravitational field $F \hat{\mathbf{e}}_x$ of the earth, the Hamiltonian for atoms trapped in the laser-field becomes
\begin{equation}\label{eq:VtrapGravitation}
H(t)=\frac{\mathbf{p}^2}{2m}
-\frac{V(t)}{1+{(z/z_0)}^2}\,
\exp\left[-\frac{2(x^2+y^2)}{w_0^2}\frac{1}{1+{(z/z_0)}^2}\right]
-F x,
\end{equation}
where $F=m g$, ($g \approx 9.81$~ms$^{-2}$, $m \approx 87$ atomic mass units for $^{87}$Rb). The depth of the optical potential $V(t)$ is freely adjustable and can be put as a function of time. If one reduces the laser power slowly, one can achieve a beam of atoms that leave the trapping potentials once it becomes unstable due to the gravitational force. Figure~\ref{fig:OpticalPotential} shows a cut trough the potential along the $x$-direction with $y=z=0$ for two different laser powers. The optical potential forms a quantum well, albeit in three dimensions, where the repulsive quadratic potential in (\ref{eq:CaseIII}) is replaced by a linear potential.

\begin{figure}[t]
\begin{center}
\includegraphics[width=0.45\textwidth]{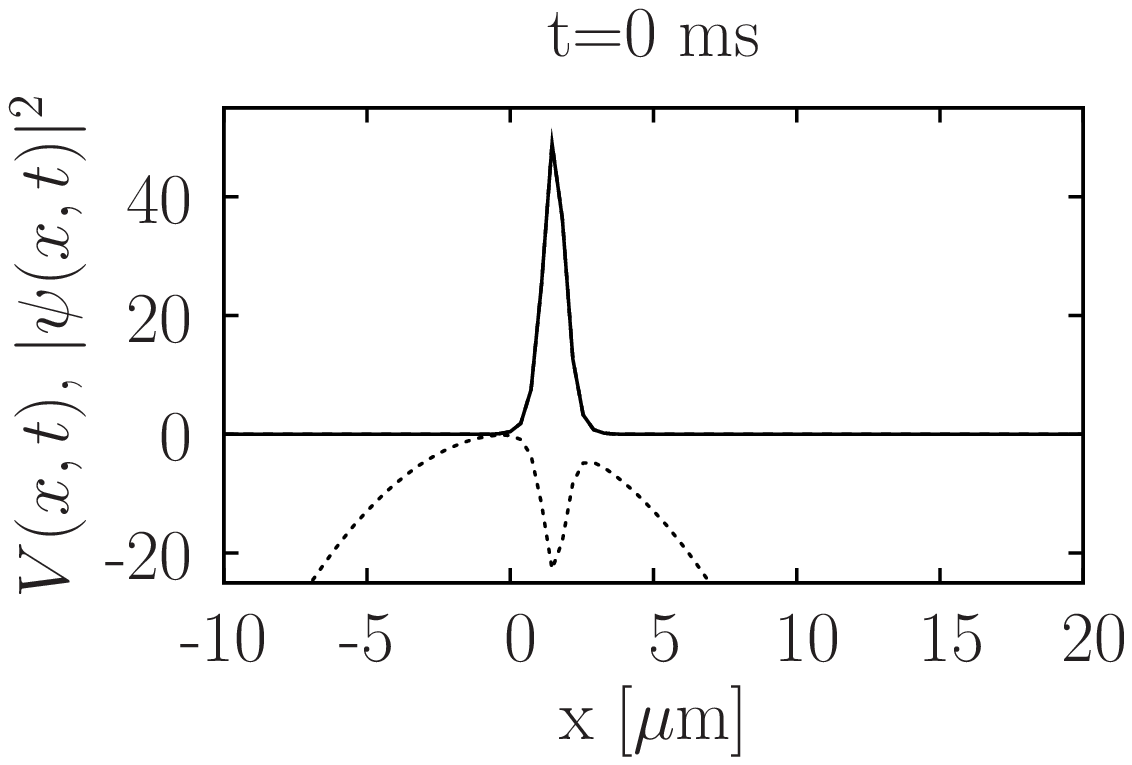}
\includegraphics[width=0.45\textwidth]{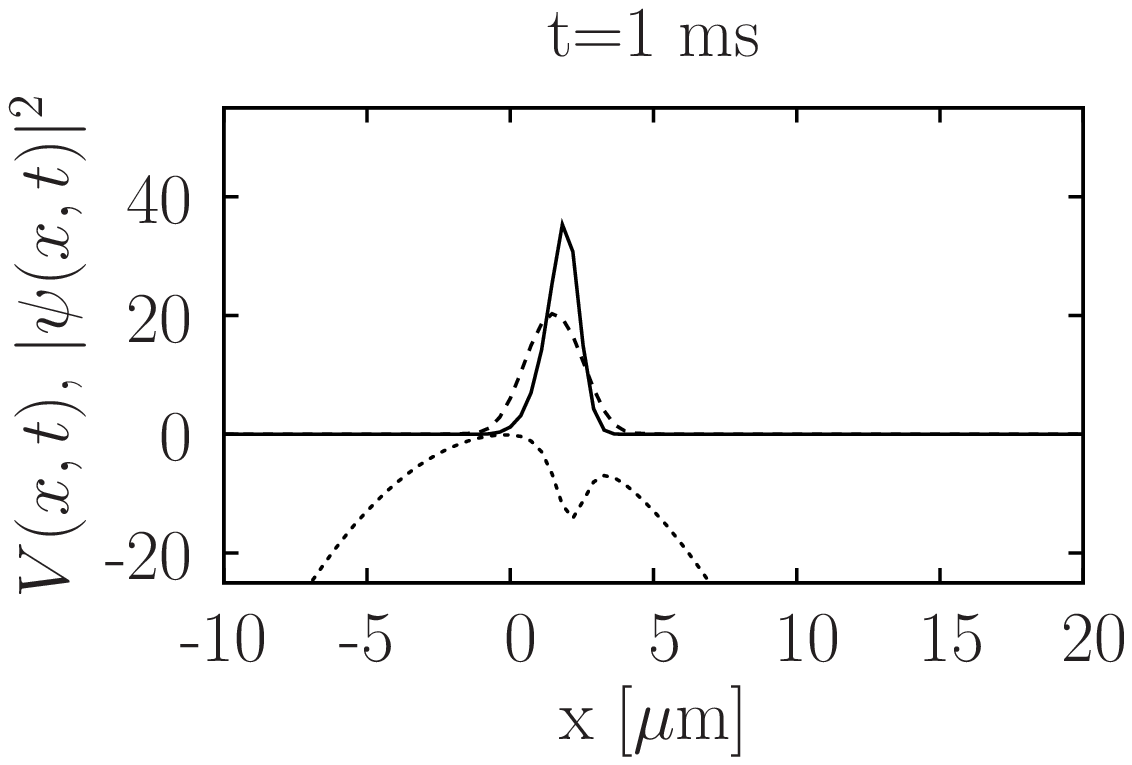}
\\
\includegraphics[width=0.45\textwidth]{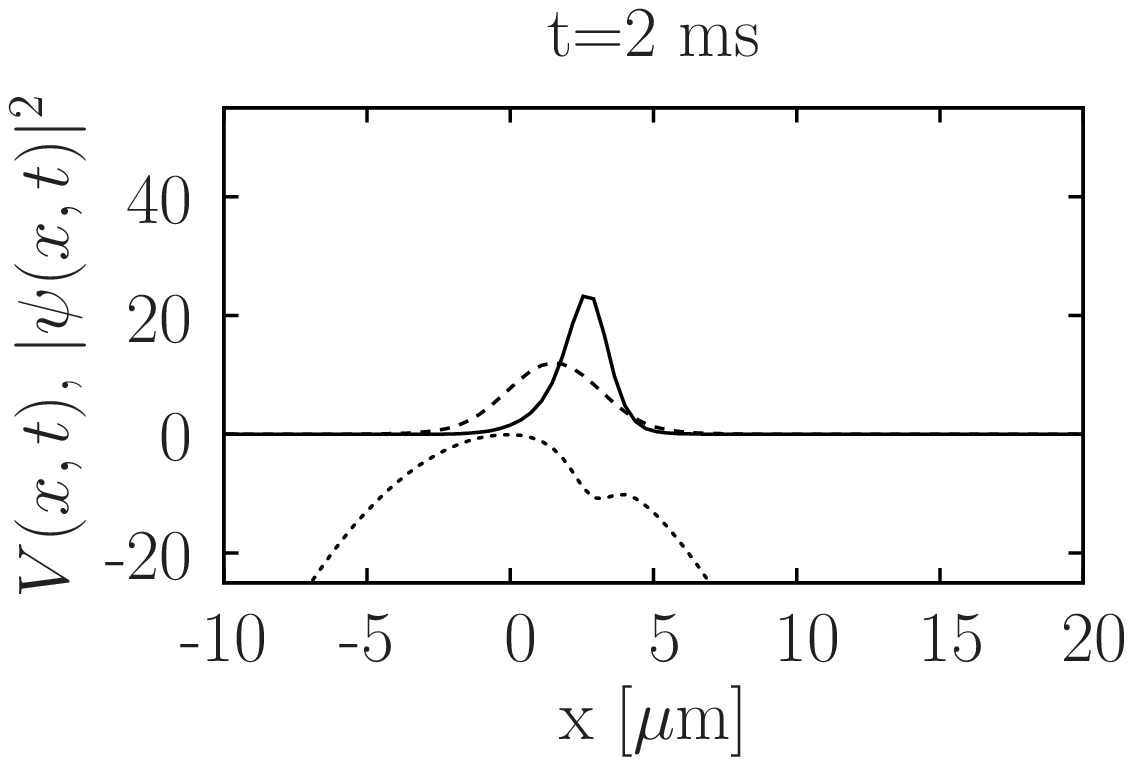}
\includegraphics[width=0.45\textwidth]{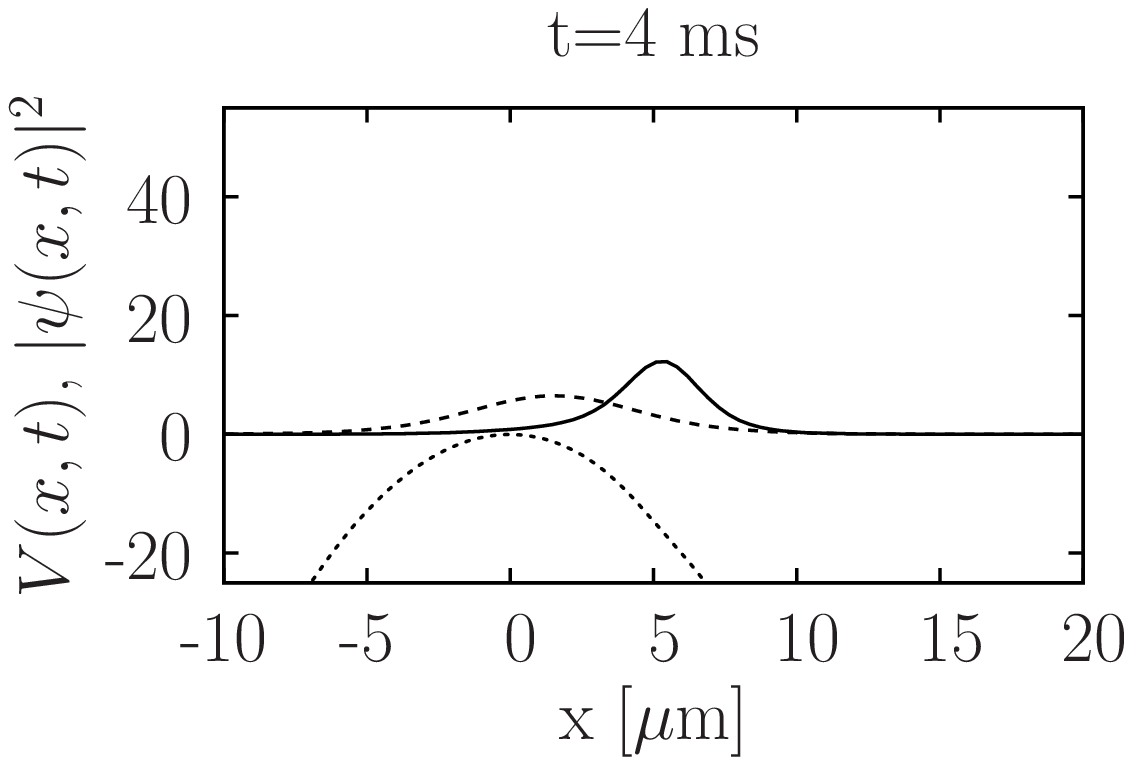}
\end{center}
\caption{\label{fig:WellDecay}
Time evolution of an initially confined wavepacket in a time-dependent potential. The dotted line shows the potential given in eq.~(\ref{eq:VchT}), the dashed line shows the force-free time-evolution of the initial state, and the solid line denotes $|\psi(x,t)^2|$ from eq.~(\ref{eq:PsiChT}).
Parameter: $\alpha=1/450$, $\mu=2\times 10^6$/m, $\omega=2\pi\times50$~Hz, the potential energy scale is $nK$, and arb.~units. for $|\psi(x,t)|^2$.}
\end{figure}

In order to model the gradually switching off of the optical trapping potential, one should take into account the asymmetry of the potential due to the gravitational force. We would also like to use the methods developed in Section~\ref{sec:Mapping}. If we choose the potential of the form
\begin{equation}
\fl
V_\ch(x)=-\frac{\alpha\mu^2\hbar^2}{m \ch^2(\alpha,\mu,x)},\quad {\rm with}\quad
\ch(\alpha,\mu,x)=\frac{1}{2}\left(\alpha\rme^{\mu x}+\rme^{-\mu x}\right),
\end{equation}
we can use the known propagator \cite{Jauslin1988a}
\begin{eqnarray}
\fl
K_\ch(x,x';t)=K_{\rm free}(x,x',t)+\frac{\alpha\mu}{2\ch(\alpha,\mu,x)\ch(\alpha,\mu,x')}
\nonumber\\
\fl
\times
\left[
\rme^{\rmi\mu^2\hbar t/(2m)}
-\left(
\rme^{\mu(x'-x)}\MoshinskyM(x-x',-\rmi\mu,t\hbar/m)+
\rme^{\mu(x-x')}\MoshinskyM(x'-x,-\rmi\mu,t\hbar/m)
\right)
\right]
\end{eqnarray}
to model a time-dependent potential similar to the one used in eq.~(\ref{eq:VDeltaISHOTime})
\begin{equation}\label{eq:VchT}
V_\ch(x,t)=-\frac{1}{2}m\omega^2 x^2+\frac{V_\ch\left(x/C(t)\right)}{C^2(t)},\quad{\rm with}\quad
C(t)=\rme^{\omega t}.
\end{equation}
A realistic (albeit one-dimensional) picture of the time-evolution of this potential is given in Fig.~\ref{fig:WellDecay}. Using the same transformation as before, we can derive the time-evolution of an initial state, which we choose in this case to be the bound eigenstate in the potential $V_\ch(x)$:
\begin{equation}
\psi_\ch(x,t=0)=\frac{\sqrt{2\alpha\mu}}{2\ch(\alpha,\mu,x)}
\end{equation}
The wave-function at a later time becomes
\begin{equation}\label{eq:PsiChT}
\fl
\psi(x,t)
=
\rme^{-\omega t/2}
\int_{-\infty}^\infty\rmd x'\;
\rme^{\rmi m \omega(x^2-x^{\prime 2})/(2\hbar)}
K_\ch\left(\rme^{-\omega t}x,\frac{1-\rme^{-2\omega t}}{2\omega}\bigg|x',0\right)
\psi_\ch(x',0).
\end{equation}
The integration has to be carried out numerically. Fig.~\ref{fig:WellDecay} shows the resulting time-evolution of the probability density for different times. The presence of the potential well delays the broadening of the initial wavepacket compared to a force-free evolution (also shown in the figure). This behaviour of the model is not desirable if one wishes to model wavepackets of longer spatial extension. The experiments so far tried to create an elongated wavepacket. To achieve this goal experimentally requires a careful tuning of the time-dependence of the trapping potential \cite{Weitz2005a}. Our simple one-dimensional model shows the coherent control of the initial wavepacket by the changing potential and may describe a more focussed atomlaser. 

A one-dimensional treatment as presented here should reproduce the main features of the experiment, since
the tunneling in the directions perpendicular to the gravitational field are not important, because the lowering of the potential is related mainly to the gravitational potential $-Fx$. For an accurate theory of the experimentally realized atomlaser \cite{Cennini2003a}, one should employ a three-dimensional trapping potential and a deeper optical potential, which can accommodate several bound states. A detailed model of an optical trap by reflectionless potentials \cite{Schonfeld1980a} is another possibility to obtain potentials which still admit analytic solutions.

\section{Conclusions}

We have conducted a systematic study of tunneling phenomena from an attractive potential emebedded into an inverted oscillator barrier. The use of the propagator makes it possible to actually obtain the complete time-evolution of the wavefunction analytically for an exponentially decaying quantum well. Our results are relevant for the study of tunneling processes in time-dependent confining potentials, such as the trapping potentials for condensed Bosons (BEC).
We succeeded in finding new analytic results for one-dimensional tunneling processes through a time-dependent barrier. The analytic results provide important reference cases for other methods, such as the semiclassical WKB approximation \cite{Maitra1997a,Saltzer2003a}. The mapping procedure of Section~\ref{sec:Mapping} in principle is not limited to one-dimensional problems and opens the way to the study the propagation of wavepackets through explicit time-dependent barriers. The experimental realization of coherent atomic ensembles in controllable confining potentials demonstrates the need for theoretical descriptions of tunneling processes through time-dependent barriers.

\ack{We appreciate stimulating discussions with M.~Kleber, P.~Kramer, V.~Man'ko, M.~Weitz, and financial support by CONACyT.
TK would like to thank the members of the Instituto de F\'{\i}sica for their hospitality during the completion of this work.}

\section*{References}

\providecommand{\url}[1]{#1}

\end{document}